\documentclass[12pt, preprint]{aastex}

\usepackage{natbib}
\usepackage{graphicx}
\usepackage{epstopdf}
\shorttitle{MOST Observations of AD Leo}
\shortauthors{Hunt-Walker et al.}
\bibpunct{(}{)}{;}{a}{}{,} 

\begin{document}

\title{{\it MOST} observations of the Flare Star AD Leo$^1$\footnotetext[1]{\lowercase{\uppercase{B}ased on observations obtained with the \uppercase{A}pache \uppercase{P}oint \uppercase{O}bservatory 3.5-meter telescope, which is owned and operated by the \uppercase{A}strophysical \uppercase{R}esearch \uppercase{C}onsortium.}}}
\author{Nicholas M. Hunt-Walker}
\email{nmhw@astro.washington.edu}
\affil{Department of Astronomy, University of Washington, Seattle, WA}
\author{Eric J. Hilton}
\email{hilton@ifa.hawaii.edu}
\affil{Department of Geology and Geophysics and Institute for Astronomy, University of Hawaii at Manoa, Honolulu, HI}
\author{Adam F. Kowalski}
\email{kowalski@astro.washington.edu}
\author{Suzanne L. Hawley}
\email{slh@astro.washington.edu}
\affil{Department of Astronomy, University of Washington, Seattle, WA}
\author{Jaymie M. Matthews}
\email{matthews@astro.ubc.ca}
\affil{Department of Physics and Astronomy, University of British Columbia, Vancouver, B.C.}
%\maketitle

\begin{abstract}
We present continuous, high-precision photometric monitoring data with 1 minute cadence of the dM3e flare star AD Leo with the {\it MOST} satellite.
We observed 19 flares in 5.8 days, and find a flare frequency distribution that is similar to previous studies.  
The light curve reveals a sinusoidal modulation with period of $2.23^{+0.36}_{-0.27}$ days that we attribute to the rotation of a stellar spot rotating into and out of view.  We see no correlation between the occurrence of flares and rotational phase, indicating that there may be many spots distributed at different longitudes, or possibly that the modulation is caused by varying surface coverage of a large polar spot that is viewed nearly pole-on.  The data show no correlation between flare energy and the time since the previous flare.  We use these results to reject a simple model in which all magnetic energy is stored in one active region and released only during flares.
\end{abstract}

%%%%%%%%%%%%%%%%%%%%%%%%%%%%%%%%%%%
%%%%%%%%%%%%%%%%%%%%%%%%%%%%%%%%%%%
\section{Introduction}
\label{sec:intro}
%%%%%%%%%%%%%%%%%%%%%%%%%%%%%%%%%%%
%%%%%%%%%%%%%%%%%%%%%%%%%%%%%%%%%%%

Flares are explosive events caused by magnetic reconnection in stellar atmospheres.
On M dwarfs, where flare signatures can be spectacular,
they emit at a range of wavelengths, from radio \citep{Osten2006} and IR \citep{Schmidt2011, Davenport2012} to
UV \citep{Audard2000, Sanz-Forcada2002,Welsh2007} and X-ray \citep{Osten2010}.
Detailed studies of flares, especially multi-wavelength observations \citep[e.g.,][]{Hawley2003,Osten2005, Smith2005}, 
allow for interpretation of the physical processes that occur during flares \cite[for a thorough review, see][]{Benz2010}.

The emission from flares can also be exploited as an indicator of the strength and distribution of the magnetically active regions on the stellar surface.
The frequency of flares of a given energy, or the flare frequency distribution (FFD), 
of a star provides one such metric by tracking how much magnetic energy is released through flares.
Photometric flare monitoring efforts \citep[e.g.,][]{Gershberg1972, Lacy1976}, typically undertaken in Johnson {\it U}, 
have found that more energetic flares occur less frequently, according to
\begin{equation}
\log \nu = \alpha + \beta \log E
\label{eqn:ffd}
\end{equation}

\noindent where $\nu$ is the frequency of flares of energy $E$ or greater, and $\beta$ is the power law exponent, typically $\lesssim -1$.

Studies of the FFD of a number of stars
have found that later-type M dwarfs flare more frequently but with less energy than their earlier-type counterparts \citep{Kowalski2010, Hilton2010, Hilton2012}.
They also found that stars with H$\alpha$ in emission, which is an indicator of magnetic activity, flare more frequently and with more energy than
inactive stars.

The FFD for the subject of this study, the bright dM3e star AD Leo, has been measured several times.
\citet{Lacy1976} found $\beta = -0.82 \pm 0.27$, although this measurement was based on only 9 flares in 21.5 hours of monitoring.
With extensive monitoring over several years, \citet{Pettersen1984} found a value of  $\beta = -0.62 \pm 0.09$, and did not see 
differences above their detection limits in the FFD from year to year, thus arguing for no cyclic variations caused by a magnetic cycle.
They also found that flares were distributed randomly in time, inferring that flares were occurring either from one spot always in view 
(nearly pole-on) or from many spots.
The former scenario is further supported by spectropolarimetric observations of \citet{Morin2008}, 
who report that the dominant feature of the surface magnetic field is likely a large polar spot, viewed nearly pole on,
with radial field of maximum magnetic flux $B = 1.3$kG.
We note that the \citet{Morin2008} result is based only on Stokes V measurements. 
Other studies using Stokes I  find larger magnetic flux values of $|{\emph B}f| \sim 3.3$ kG \citep{Johns-Krull2000}, $|{\emph B}f| = 2.9$ kG \citep{Reiners2007}, and $|{\emph B}| = 2-2.5$ \citep{Shulyak2010}.

A new generation of space missions allows for previously inaccessible continuous, short cadence, high precision photometric monitoring of flares.
These satellites have sufficient precision to measure rotational modulation caused by star spots \citep[e.g.,][]{Rucinski2004,Basri2011, Harrison2012} and to detect small flares.
Practically continuous monitoring ensures that nearly all flares are observed.
For the first time, we can use the relative timing of flare occurrence to test simple models of the distribution of the active regions on M dwarfs.

We present the results of 8 days of continuous photometric monitoring observations of AD Leo with the \emph{Microvariability and Oscillations of STars} (\emph{MOST}) satellite. 
The details of the observations and photometry are provided in Section \ref{sec:photo},
along with our methods for processing the light curve and for identifying flares.
Our measurements of the FFD and an analysis of the timing of the observed flares  are described in Section \ref{sec:results},
where we compare these results to a simple, single active region model.
We conclude in Section \ref{sec:conclusion}.

%%%%%%%%%%%%%%%%%%%%%%%%%%%%%%%%%%%
%%%%%%%%%%%%%%%%%%%%%%%%%%%%%%%%%%%
\section{{\it MOST} Photometry}
\label{sec:photo}
%%%%%%%%%%%%%%%%%%%%%%%%%%%%%%%%%%%
%%%%%%%%%%%%%%%%%%%%%%%%%%%%%%%%%%%

The {\it MOST} mission is a Canadian microsatellite tasked with the detection and characterization 
of acoustic oscillations in Sun-like stars, metal-poor subdwarfs, and magnetic stars in order to seismically probe their structures and ages.  
{\it MOST} travels in a sun-synchronous, polar orbit with a period of 101.4 minutes, allowing it to observe a single target within the Continuous Viewing Zone 
for up to $\sim$60 days at a time without interruption \citep{Rowe2006a}.  
The sensitivity and speed of {\it MOST} provides micromagnitude precision for stars with $0.4<V<5.5$ with individual exposure times of less than $1$ minute \citep{Walker2003}.
The less precise direct imaging mode, which was used for the data presented here (all targets in the range $6.5<V<13$), achieves precision of $\sim$0.1 millimags \citep{Rowe2006}.

Photometric observations of AD Leo ($V=9.4$) were obtained with {\it MOST} between 6-14 March 2010,
which corresponds to 3717-3726.5 in the {\it MOST} standard system of JD-J2000 (JD - 2451545, the epoch of January 1, 2000).
Our observations have a cadence of 60 seconds through a broad optical bandpass of $\sim$3500-7000\AA.  
Guide stars were captured in 3-second exposures and stacked in groups of 10.  

Photometry was performed by the {\it MOST} pipeline \citep[described in detail in][]{Rowe2006}.
Several of the processing steps that are typically performed by the full pipeline, namely those that clean the light curve and attempt to remove spurious signals and outlier points, 
were not applied to the AD Leo data since they carried the risk of filtering out legitimate flare signals.
The partially-processed light curve will hereafter be referred to as the ``raw'' light curve. 

\begin{figure}[t]
\includegraphics[width=170mm]{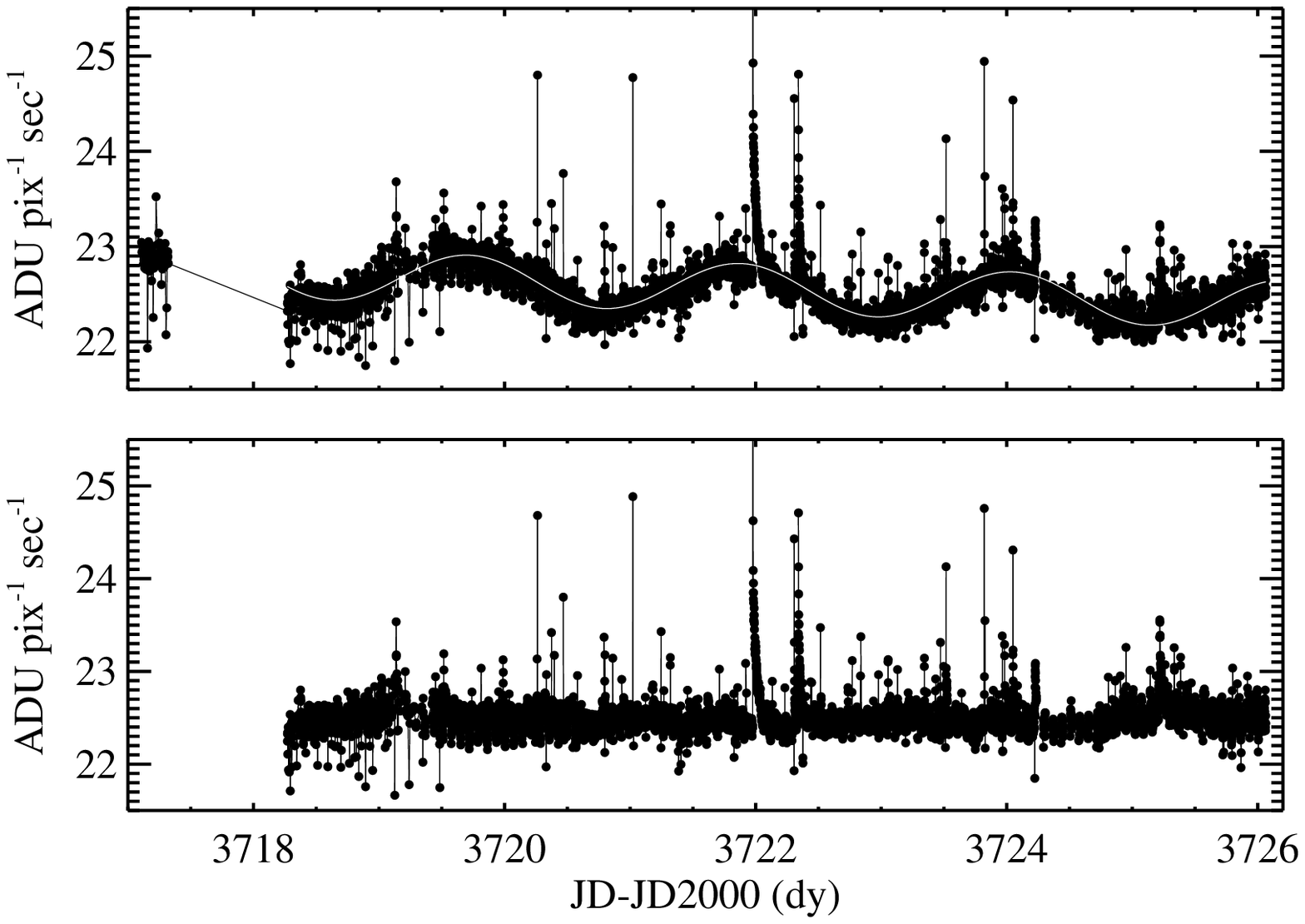}
\caption{\emph{Top:} The raw {\it MOST} light curve of AD Leo.  The gray sinusoid is the calculated rotational period.  
\emph{Bottom:} Residual model-subtracted light curve with stray light modulation and flare candidates.  Epochs prior to 3718.27 JD-JD2000 dy have been removed due to insufficient observations.}
\label{fig:both_lc}
\end{figure}

The \emph{MOST} observations, shown in the top panel of Figure \ref{fig:both_lc}, produced a time series of 8,592 epochs over 9 days.
The first 89 epochs, representing only 1\% of the total, were followed by a 23-hour gap and were neglected from further analysis.
Figure \ref{fig:both_lc} (bottom) shows the raw time series after partial {\it MOST} pipeline processing and after a sinusoid was removed
 (described further in Section \ref{sec:period}).  
%The flux measurements remain in instrumental flux units of ADU pix$^{-1}$ sec$^{-1}$. 

\begin{figure}[t]
\includegraphics[width=180mm]{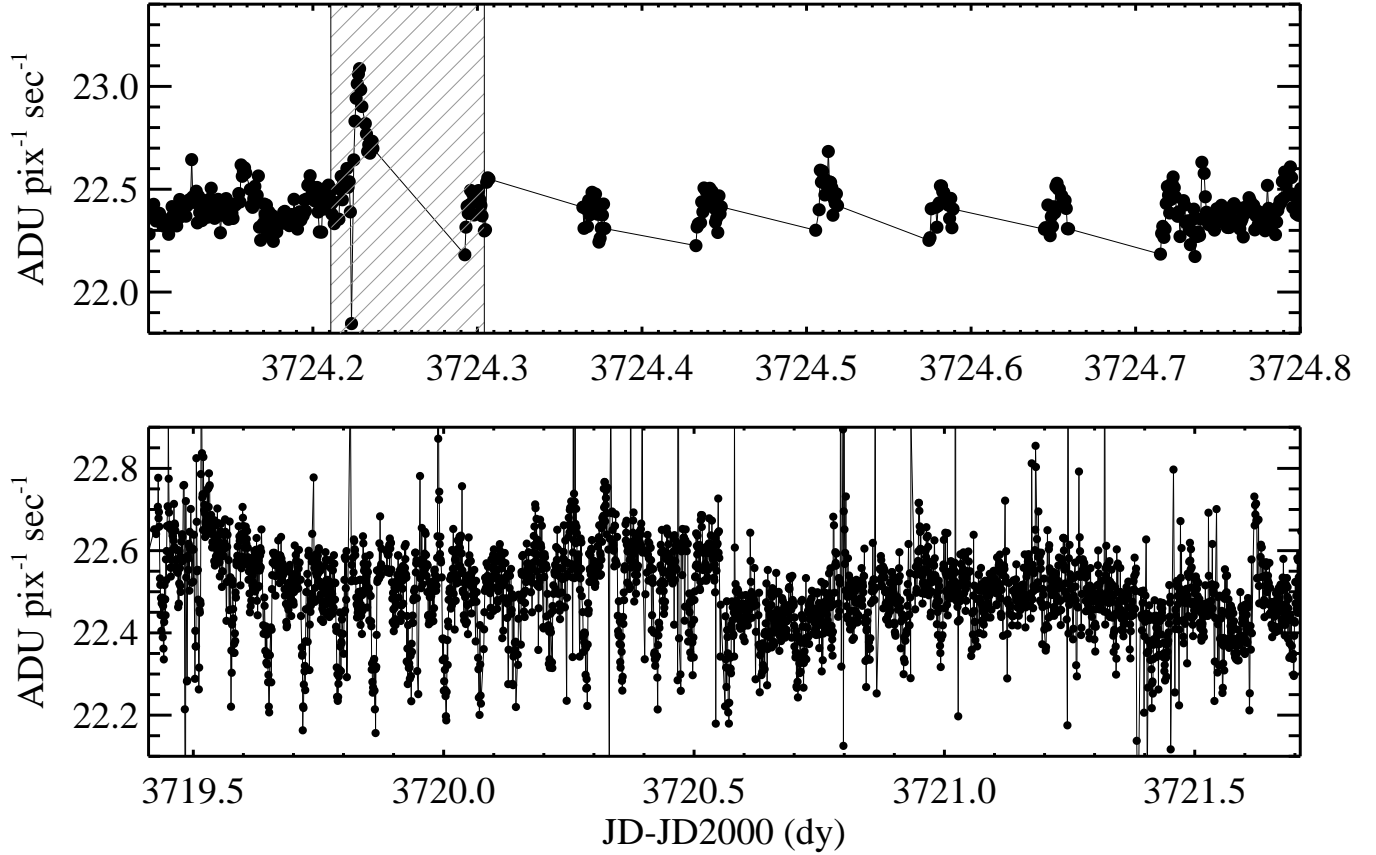}
\caption{\emph{Top:} An enlarged section of the light curve, showing the truncated flare between 3724.2 and 3724.4 dy (\emph{hatched region}).  
This flare is not counted amongst the final flare candidates due to insufficient data.  \emph{Bottom:} Modulation of the light curve due to stray light, from 2.3 - 4.6 days after 3717.1084 JD-JD2000 dy.  
The period of this modulation corresponds to the 101.4 minute orbital period of {\it MOST}.  Epochs of dimming due to this modulation were masked when calculating the FIA.}
\label{fig:badlcstuff}
\end{figure}

Many exposures were missing from the raw light curve due to instrumental issues or suspected cosmic rays.
These missing exposures are most noticeable near JD-JD2000 = 3724.5, which is also shown in more detail in Figure \ref{fig:badlcstuff} (top).
The gaps in this period prevent us from successfully identifying and measuring all flares.
We therefore additionally neglect the data in between JD-JD2000 = 3724.2364 and 3724.7151.
Additional short gaps longer than three minutes are also removed from the overall monitoring time.
Our final light curve contains 8383 points, corresponding to a total of 5.82 days of observations.

The {\it MOST} satellite experiences phases of increased stray light either from the illuminated side of the Earth \citep{Walker2003} 
or the moon that occur regularly during each pass of its orbit \citep{Rowe2006}. 
These phases last for 43.2 minute intervals at a period of 101.4 minutes, and produce non-homogeneous intensity variations.
Typically, this effect is removed from {\it MOST} lightcurves by folding the time series with the satellite's orbital period, 
calculating a running average, and correcting the entirety of the time series by this average.
However, because the amplitude of the stray light modulation varies over time, there are over- or under-compensations in different parts of the time series,
some of which can mimic a flare signal.
Figure \ref{fig:badlcstuff} (bottom) shows the most prominent of these fluctuations in the time series, between 2.3 and 4.6 days after the start of observations.  
The dimming ranges from $\sim$0.4 ADU pix$^{-1}$ sec$^{-1}$  below the median, to less than the rms scatter ($\le$0.2 ADU pix$^{-1}$ sec$^{-1}$).
We therefore did not correct the lightcurve and instead account for the periodic dimming signature when searching for flares, as we discuss in detail in Section \ref{sec:FIA}.

%The removal of this interference was attempted by reduction of the data within the {\it MOST} pipeline.  
%The reduction procedure involves folding the time series with the satellite's orbital period, calculating a running average, and correcting the entirety of the time series by this average.  
%However, because the amplitude of the stray light modulation varies over time, there is an over- or under-compensation in different parts of the time series.  
%Drastic periods of overcompensation produce the prominent dimming seen in Figure \ref{fig:stray_light}.  
%This dimming ranges from $\sim$0.4 ADU pix$^{-1}$ sec$^{-1}$  below the median, to less than the rms scatter ($\le$0.2 ADU pix$^{-1}$ sec$^{-1}$ ) of the adjusted time series.  
%There doesn't seem to be any apparent undercompensation in the light curve.  
%This fluctuation in the time series is discussed further in Section \ref{sec:FIA}

%%%%%%%%%%%%%%%%%%%%%%%%%%%%%%%%%%%
\subsection{Modeling and Removal of the Periodicity of the Light Curve}	
\label{sec:period}
%%%%%%%%%%%%%%%%%%%%%%%%%%%%%%%%%%%

\begin{figure}[t]
\includegraphics[width=180mm]{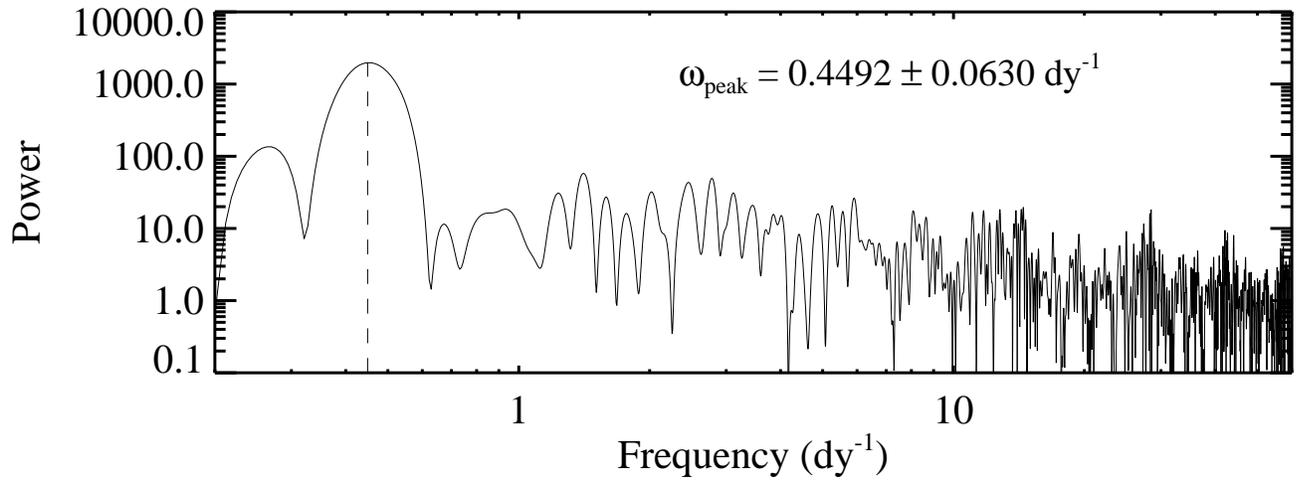}
\caption{Results of the Lomb-normalized periodogram showing the power corresponding to the
peak power of the periodogram at a rotational frequency of $\omega = 0.449\pm0.063$ dy$^{-1}$ (dashed line). 
This period agrees well with the 2.2399 dy ($\omega =$ 0.4464 dy$^{-1}$) period of \citet{Morin2008}.
The orbital period of the spacecraft ($\omega = 14.2 $dy${-1}$) does not produce a significant peak.
}
\label{fig:periodogram}
\end{figure}

The raw {\it MOST} light curve of AD Leo was dominated by a large, regular modulation that 
we interpret to be caused by a star spot rotating into and out of view, as in \citet{Pettersen1984}.  
The frequency of the modulation, $\omega$, was obtained using a Lomb-normalized periodogram and is shown in Figure \ref{fig:periodogram}.  
We note that the orbital period of the spacecraft ($\omega = 14.2 $dy$^{-1}$) does not produce a noticeable peak.
%The uncertainty in this frequency was initially estimated using the prescription from \citet{Kovacs1981}:  
%
%\begin{equation}
%\delta\omega = \frac{3\pi\sigma_N}{2(N_0)^{1/2}TA},
%\end{equation}
%
%where $\sigma_N^2$ is the variance of the noise after the signal was subtracted, $A$ is the amplitude of the signal, 
%$N_0$ is the number of elements in the data set, and $T$ is the total temporal length of the data set.  
%The uncertainty produced of 6.278$\times10^{-7}$ dy$^{-1}$, however, was deemed too small to be reasonable.  
We adopt the FWHM of a Gaussian fit to the periodogram peak as the uncertainty in the period.
The frequency of 0.449$\pm0.063$ dy$^{-1}$ corresponds to a rotational period of 2.23$^{+0.36}_{-0.27}$ dy,
in good agreement with the 2.24 dy period found by \citet{Morin2008}.
%The periodogram shows no evidence for a period of 2.7 days, as found by \citet{Spiesman1986}.

Identifying flares in the light curve is best accomplished when any signatures not caused by flaring have been removed.
An empirical model, $F(t)$, of the form 
%, the modulation of which is presumed to approximately be due to the rotation of a star spot on the surface of AD Leo (reference would help here):
\begin{equation}
F(t) = \frac{A}{2}\sin[2\pi \omega(t - B)] + Ct + D
\end{equation}
was fit to and then subtracted from the light curve.
The parameters A through D (0.5161, 0.9007, 0.0406, 0.2053) were determined by a least-squares fit of the model to the light curve, using the frequency ($\omega$) determined above.
%The light curve was set to a zero-point median by subtracting the median flux (22.504388 ADU pix$^{-1}$ sec$^{-1}$) from every point.
%\begin{eqnarray}
%F = \frac{0.5161437}{2}\sin[2\pi (\omega)(t - 0.90074855)] - (0.04056275)t + 0.20531682
%\end{eqnarray}

%%\begin{deluxetable}{ccccc}[h]
%\begin{deluxetable}{cc}
%\label{tab:sinusoid_fit}
%\tablewidth{0pt}
%\tablenum{2.1}
%\tablecaption{Model Fit}
%\tabletypesize{\scriptsize}
%\tablehead{
%\colhead{Parameter}  & Value \\}
%%\colhead{Parameter}  & \colhead{A} & \colhead{B} & \colhead{C} & \colhead{D} \\ }
%\startdata
%A & 0.5161 \\
%B & 0.9007 \\
%C & 0.0406 \\
%D & 0.2053 \\
%%Value & 0.5161 & 0.9007 & 0.0406 & 0.2053\\
%\enddata
%\end{deluxetable}

The linear term accounts for the small observed decrease in flux over the course of the observations
that may be caused by a much longer time-scale modulation or by instrumental effects.  
In either case, it affects neither our period determination nor our flare identification and can therefore safely be removed.  

Figure \ref{fig:both_lc} (bottom) shows the resulting flattened light curve with the empirical model removed.  
The small-scale structure remaining within the light curve is due to systematic noise 
(such as stray light from the illuminated side of the Earth -- Section \ref{sec:photo}) and intrinsic stellar variability, such as candidate flares.

%%%%%%%%%%%%%%%%%%%%%%%%%%%%%%%%%%%
\subsection{The Flare Identification Algorithm}
\label{sec:FIA}
%%%%%%%%%%%%%%%%%%%%%%%%%%%%%%%%%%%

Flare candidates were identified using the Flare Identification Algorithm (FIA) of \citet{Hilton2011, Hilton2012}.
Briefly, at each epoch, $i$, the running local median and standard deviation ($\sigma_{local}$) are computed using the 31 epochs nearest in time to $i$ that 
have not been flagged as candidate flares.
The procedure consists of iteratively identifying consecutive epochs that fall sufficiently far above the local median, flagging them as flare candidates, 
masking them so that they are not included in the determination of the local standard deviation, 
and repeating those steps on the masked light curve until new candidates are no longer found.  
The criteria that were used to identify flare candidates required $\ge$ 3 epochs above 2.5$\sigma_{local}$, at least one of which is above 3.5$\sigma_{local}$.
Flare candidate start and stop times were determined by identifying the last epoch prior to -- and the first epoch after-- the flare peak in which the running mean of 5 epochs was less than 0.5$\sigma_{local}$.

As discussed in \S \ref{sec:photo}, some portions of the light curve contained periodic decreases in flux for which corrections were not applied.
The FIA computes a running standard deviation that would become larger in the sections with decreased flux, thereby increasing the minimum flux threshold that the light curve
would have to reach to be considered a flare.
Because the increase in $\sigma_{local}$ is caused by systematic effects and not a decrease in the photometric precision, during the periods 
when the dimming occurs, we mask the dimmed regions in our computation of the running local median and standard deviation, in effect treating them the same as the flare epochs that are
also removed.
Although this masking allows us to detect flares near these regions, flares that occur during or near these periods must be slightly larger than other, smaller flares in order to be detected.
For the light curve, the difference between the median values during the orbital dimming periods and outside of the orbital periods is less than 0.01 ADU pix$^{-1}$ sec$^{-1}$ .
For the period shown in the bottom panel of Figure \ref{fig:badlcstuff}, the total time when the light curve is more than 0.1 ADU pix$^{-1}$ sec$^{-1}$ below the median is 2.6 hrs.
This is less than 2\% of the total monitoring time and has negligible effect on the final FFD.

\begin{figure}[t]
\includegraphics[width=170mm]{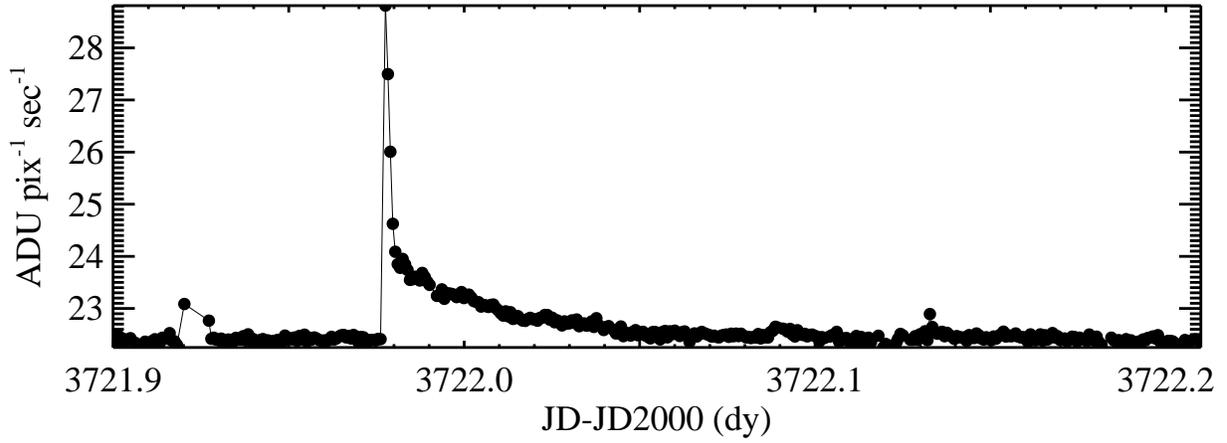}
\caption{An enlarged section of the light curve showing the largest flare candidate in the time series.  
This flare lasts for 3.97 hours, beginning at 3721.97 dy and returning to quiescence at 3722.05 dy.  
The peak of this flare is at 28.83 ADU pix$^{-1}$ sec$^{-1}$ representing an increase of $28\%$ over the average quiescent value of the time series.}
\label{fig:big_flare}
\end{figure}

Once the FIA was applied, the results were visually inspected and
the flare candidates that did not resemble the classical flare shape of a sharp rise with an exponential decay were rejected.  
The classic flare is shown in Figure \ref{fig:big_flare}.
In one case, shown in Figure \ref{fig:badlcstuff} (top), a candidate flare with a sharp rise and a portion of the initial decay phase occurred at the beginning of a period that was neglected because of missing data.
We do not include this flare in the majority of our analysis, although we we will mention it briefly when we consider the flare wait time distribution (see Section \ref{sec:flare_timing}).

%%%%%%%%%%%%%%%%%%%%%%%%%%%%%%%%%%%
%%%%%%%%%%%%%%%%%%%%%%%%%%%%%%%%%%%
\section{Results}
\label{sec:results}
%%%%%%%%%%%%%%%%%%%%%%%%%%%%%%%%%%%
%%%%%%%%%%%%%%%%%%%%%%%%%%%%%%%%%%%

\begin{figure}[t]
\includegraphics[width=160mm]{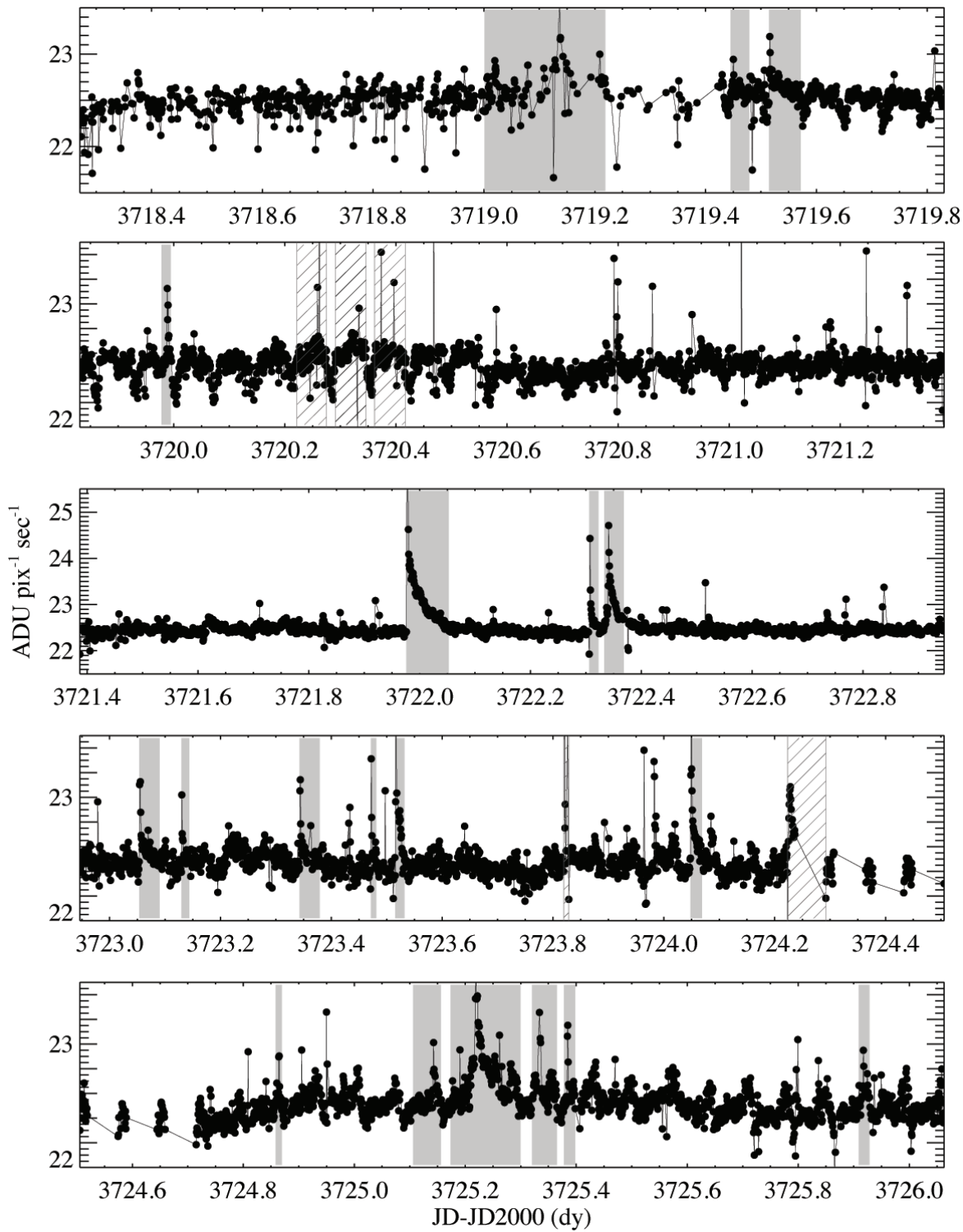}
\caption{The model-subtracted light curve with stray light modulation.  
Solid shaded regions identify the 19 flares that passed visual inspection out of the original 24 candidates.  
The rejected flare candidates are shown as hatched regions.}
\label{fig:full_lc}
\end{figure}

When applied to the \emph{MOST} time series, the FIA resulted in 24 flare candidates, of which 19 passed visual inspection.  
These flare candidates are shown in shaded regions in Figure \ref{fig:full_lc}, while the detected but rejected flares are 
shown as hatched regions.
One of the rejected flares is the truncated flare shown in Figure \ref{fig:badlcstuff} (top) that occurred during a neglected portion of the light curve.
The other rejected candidates occurred during the continuous portion of the light curve, 
but did not show the typical profile of stellar flares. 
They may comprise a series of short, complex flares that were not well-resolved, or they may be epochs that are compromised by data loss, poor photometry, cosmic rays, etc.

Of the 19 confirmed flares, the largest (Figure \ref{fig:big_flare}) reaches its maximum at 6 ADU pix$^{-1}$ sec$^{-1}$ above the median and lasts for 3.97 hrs.
The peak was over 200 times the $\sigma_{local}$ value, with a total energy release of $log E = 33.5$ ergs.
The smallest flare was roughly 0.4 ADU pix$^{-1}$ sec$^{-1}$ above the mean, lasting for 0.12 hrs ($\sim$7 minutes).  
The peak was 10 times the $\sigma_{local}$ value, with a total energy release of $log E = 31.6$ ergs.

To compute the flare energy, we first 
convert the instrumental ADUs to physical units through spectrophotometry.
We use a flux-calibrated spectrum of AD Leo from \citet{Kowalski2012} obtained with the Dual-Imaging Spectrograph on the ARC 3.5m telescope at Apache Point Observatory, 
and convolve it with the {\it MOST} bandpass.
We then calculate the dimensionless flux from the flare, normalized by the quiescent flux of the star, as 
\begin{equation}
F_{f}(t) = \frac{F_{Flare}(t) + F_0}{F_0} - 1 = \frac{F_{Flare}(t)}{F_0}
\label{eqn:flare_intensity}
\end{equation}
where $F_0$ is the quiescent flux.  We then compute the equivalent duration, $P$ \citep{Gershberg1972}, which is defined as the amount of time that it would take the star, in its quiescent state, 
to release the same amount of energy released during the flare and is simply the time integral of $F_f$.
\begin{equation}
P = \int \! F_{f}(t) \, dt
\label{eqn:equiv_duration}
\end{equation} 

%The typical method \citep[e.g.,][]{Lacy1976, Pettersen1984,Walkowicz2011, Hilton2012} for computing flare energies is to multiply 
The flare energy is the quiescent luminosity of the star, $q$, multiplied by the equivalent duration of the flare, $P$:
\begin{equation}
E_{flare} = q \times P
\end{equation}.

\noindent We stress that these values are filter dependent, such that flare energies measured in different filters are not necessarily equal.
Finally, the cumulative FFD is found from the ordered list of flare equivalent durations, or $P_i$ for the $i$th flare.
The frequency of a flare with $P > P_i$ is $i$ (which is the number of flares larger than $P_i$) divided by the effective monitoring time.
Because the {\it MOST} photometry has similar precision throughout the light curve, $i$ is essentially divided by the cumulative amount of time where there are observations.   
Note that the total monitoring time does account for gaps in the otherwise continuous observations.

% Adam sez: The filter-weighted flux (at earth) of AD Leo that I calculate is 8.1e-13 ergs/s/cm^2/Ang
% FWHM of most bandpass is 298.98075 nm
% Adam sez: "To calculate the MOST luminosity just multiply by 4pi x dist^2 and the filter bandpass full width at half max."
% 4.9 pc = 1.512 x 10^{19} cm
% Resulting quiescent luminosity: 6.957 x 10^{30}

We find that $q_{MOST} = 6.96\times10^{30}$ erg s$^{-1}$ using the flux determinations described above and a distance of 4.9 pc for AD Leo \citep{Perryman1997}.
However, because the {\it MOST} photometry is dominated by red light, and previous flare studies have used a narrower and bluer filter, 
such as Johnson {\it U}, the flare energies cannot be directly compared.  
%Instead, we use the equivalent duration of the flare as a proxy for flare energy, noting that the two values differ only by a multiplicative constant.
Therefore we compare the frequencies, normalizing the equivalent durations to the largest value of $\log P$, allowing us to compare the values of $\beta$ from Equation \ref{eqn:ffd}.
The flare energies are used for the analysis of flare timing (see Section \ref{sec:flare_timing}).
%This converts Equation \ref{eqn:ffd} into 
%$\log \nu \propto \beta \log P$.

\begin{figure}[t]
\includegraphics[width=170mm]{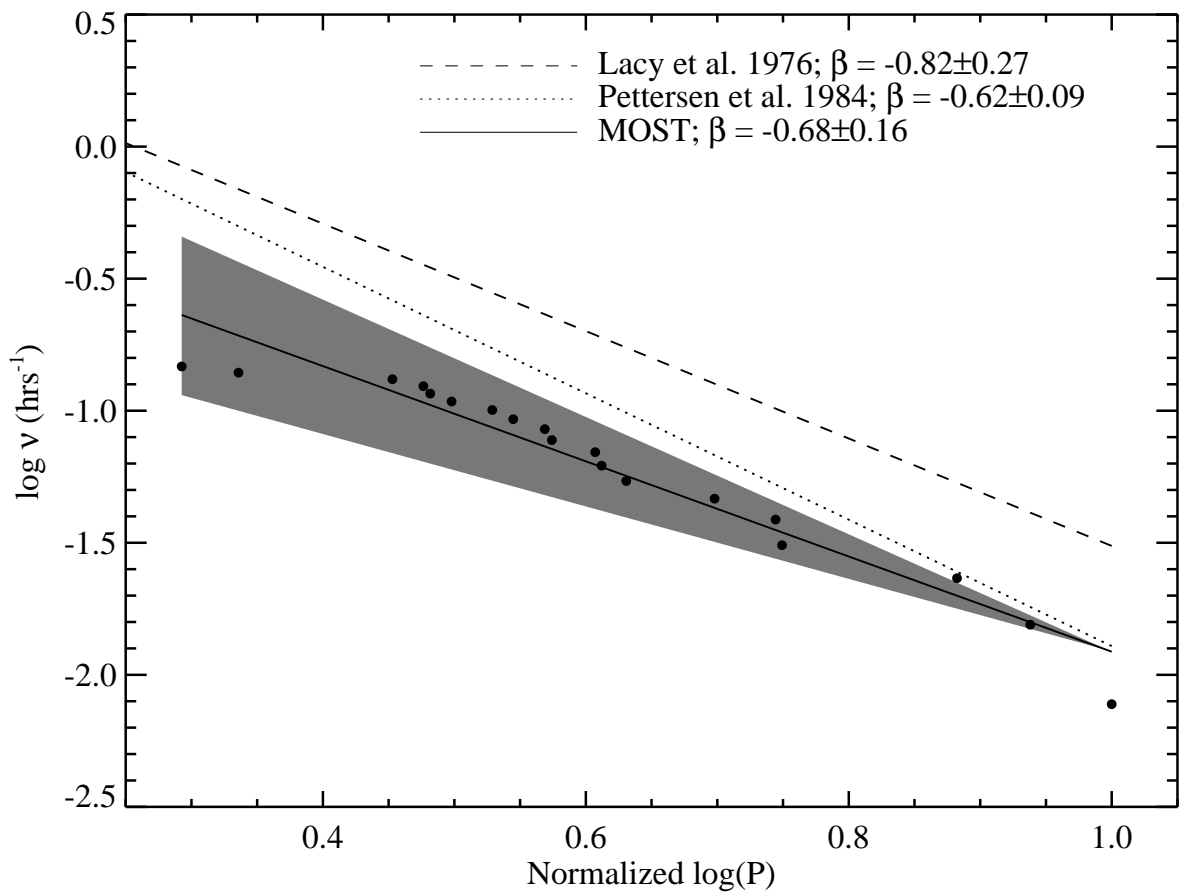}
\caption{Cumulative flare frequency distributions (FFD), normalized to the maximum equivalent durations ($P$) for each distribution.  
The FFDs have been normalized to facilitate comparison between the observations take in the {\it MOST} filter and those taken in the {\it U} filter.
The {\it MOST} observations (solid line) show less frequent flaring, but have a similar $\beta$ value as the FFDs from \citet[][dotted line]{Pettersen1984}
and \citet[][dashed line]{Lacy1976}.}
The uncertainty in $\beta$ for the {\it MOST} observations is shown as the shaded region.
\label{fig:ffd}
\end{figure}

The normalized FFD from our observations is shown in Figure \ref{fig:ffd} (black line and points), along with the normalized FFDs for AD Leo from \citet[][115 flares]{Pettersen1984}
 and \citet[][9 flares]{Lacy1976}. 
It should be noted that the fits are weighted towards high-energy flares, 
since all of these studies are affected by detection thresholds at the low energy end.
We find $\beta = -0.68\pm0.16$, which is in agreement with \citet[][$\beta = -0.82\pm0.27$]{Lacy1976} and \citet[][$\beta = -0.62\pm0.09$]{Pettersen1984}.
We compute uncertainties in the same manner as these previous studies, with $\sigma_\beta = \frac{\beta}{\sqrt{n}}$, where $n$ is the number of flares.
The general agreement is remarkable, especially considering that the other studies
were performed in the Johnson {\it U} filter.
Flares release more energy at bluer wavelengths \citep[e.g.,][]{Hawley1991, Kowalski2010}, and AD Leo has much less quiescent flux in {\it U} than in the broader
{\it MOST} filter.
The broad filter means that despite the sensitivity of {\it MOST}, only relatively large flares are detected, and the overall flare rate is lower than it is in the {\it U} filter.

\subsection{Flare Timing}
\label{sec:flare_timing}

The {\it MOST} light curve is nearly continuous over several days, with enough precision to detect photometric variations on timescales of a day or longer. 
We are thus able to investigate when flares occur relative to other flares and relative to the (presumed) rotational phase.
These studies cannot be easily done from the ground due to diurnal observing interruptions.
    
We wish to use the timing and energies of flares as probes of the distribution of magnetically active regions on the stellar surface.
In this paper, we adopt a simple model that can be tested with the observations.
We assume that all of the magnetic energy that is generated by the stellar dynamo is stored in active regions during quiescence, 
and is only released during flares. 
We also assume a single, large active region which stores the magnetic energy.
This assumption is supported by the strong periodic signature in the {\it MOST} light curve, 
which we interpret as the rotational modulation of a large active region. 
It is further supported by the spectropolarimetric observations of \citet{Morin2008}, who report that one large polar region, viewed nearly pole-on,
is the dominant magnetic feature on AD Leo.
In our model, then, the wait time between a flare and the proceeding flare should be proportional to the flare energy.
Furthermore, flares should be more likely to be observed when the active region is visible, which corresponds to the troughs of the light curve assuming that the active region is dark in the \emph{MOST} filter.

\begin{figure}[t]
\includegraphics[width=180mm]{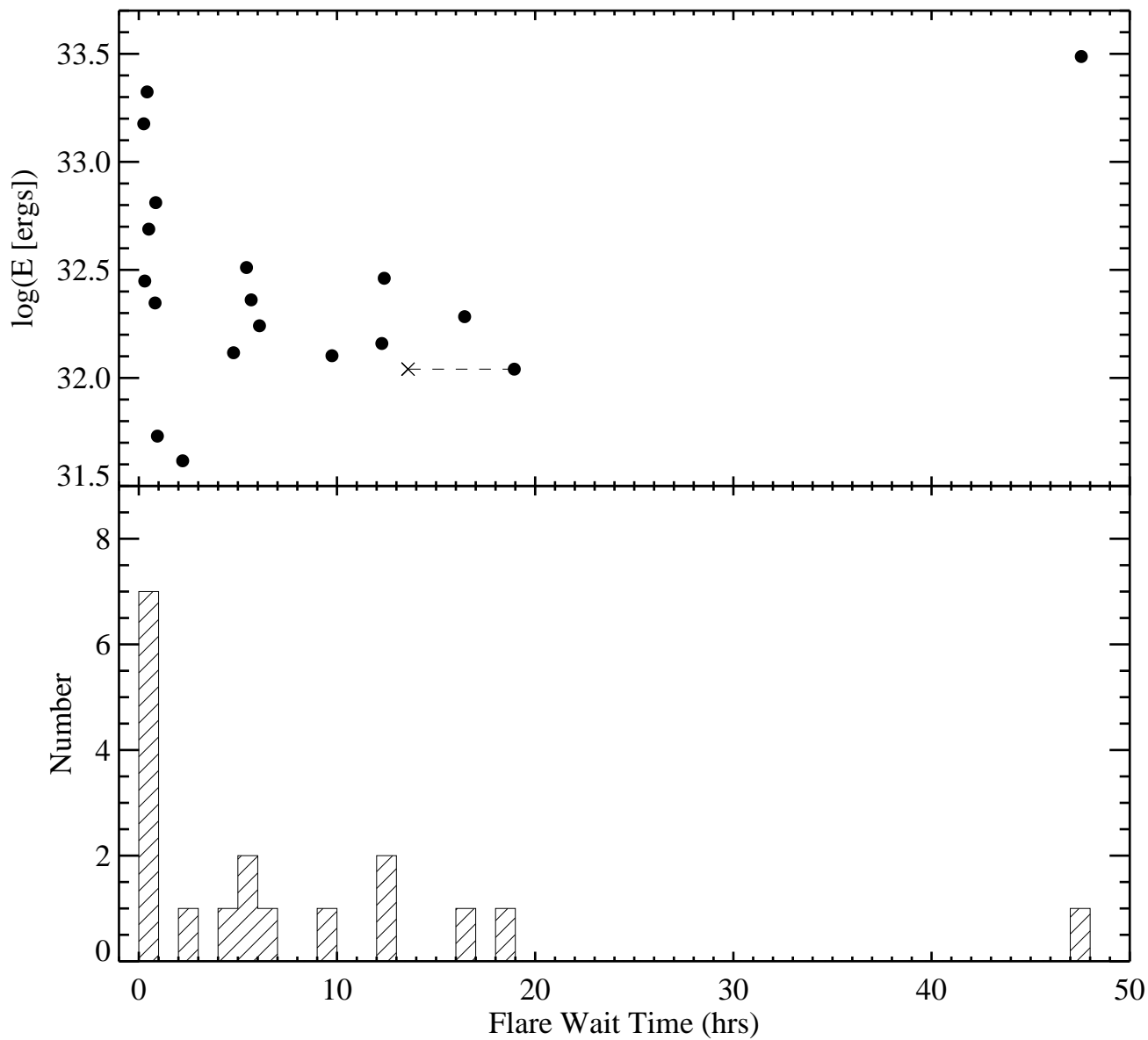}
\caption{\emph{Top:} Flare energy is not strongly correlated with the time since the previous flare.
Although the largest flare in the sample also has the longest wait time, there are several large flares with very short wait times.
The second longest wait time occurred after the truncated and rejected flare in Figure \ref{fig:badlcstuff} (top). 
Including the rejected flare reduces the wait time to $\sim$12 hours, as shown by the dashed line.
\emph{Bottom:} 
The distribution of wait times between flares in bins of 1 hour, shows that most flares occur within a few hours of the previous flare.}
\label{fig:delay_distribution}
\end{figure}

We plot the flare energies against the time since the previous flare in the top panel of Figure \ref{fig:delay_distribution}.
The bottom panel displays the number of flares in each 1 hour bin of wait time.
 The longest wait time between flares does occur prior to the largest flare in the set.  
The next longest delay is 18.95 hrs, leading to the flare candidate beginning at 3724.86 dy.  
While this flare candidate is not notably large, it must be noted that this is the first flare to occur after the neglected portion of the light curve.
At least one flare appears to have occurred during this period (the truncated flare shown in Figure \ref{fig:badlcstuff}, top).  
If we consider this to be the most recent flare, the delay of 18.95 hrs is closer to 13.59 hrs, which is marked by the dashed line.  
In any case, the data in Figure \ref{fig:delay_distribution} do not show a strong correlation between wait time and flare energy.  In fact, some of the largest flares in the sample occur after very short delay times.

\begin{figure}[t]
\includegraphics[width=180mm]{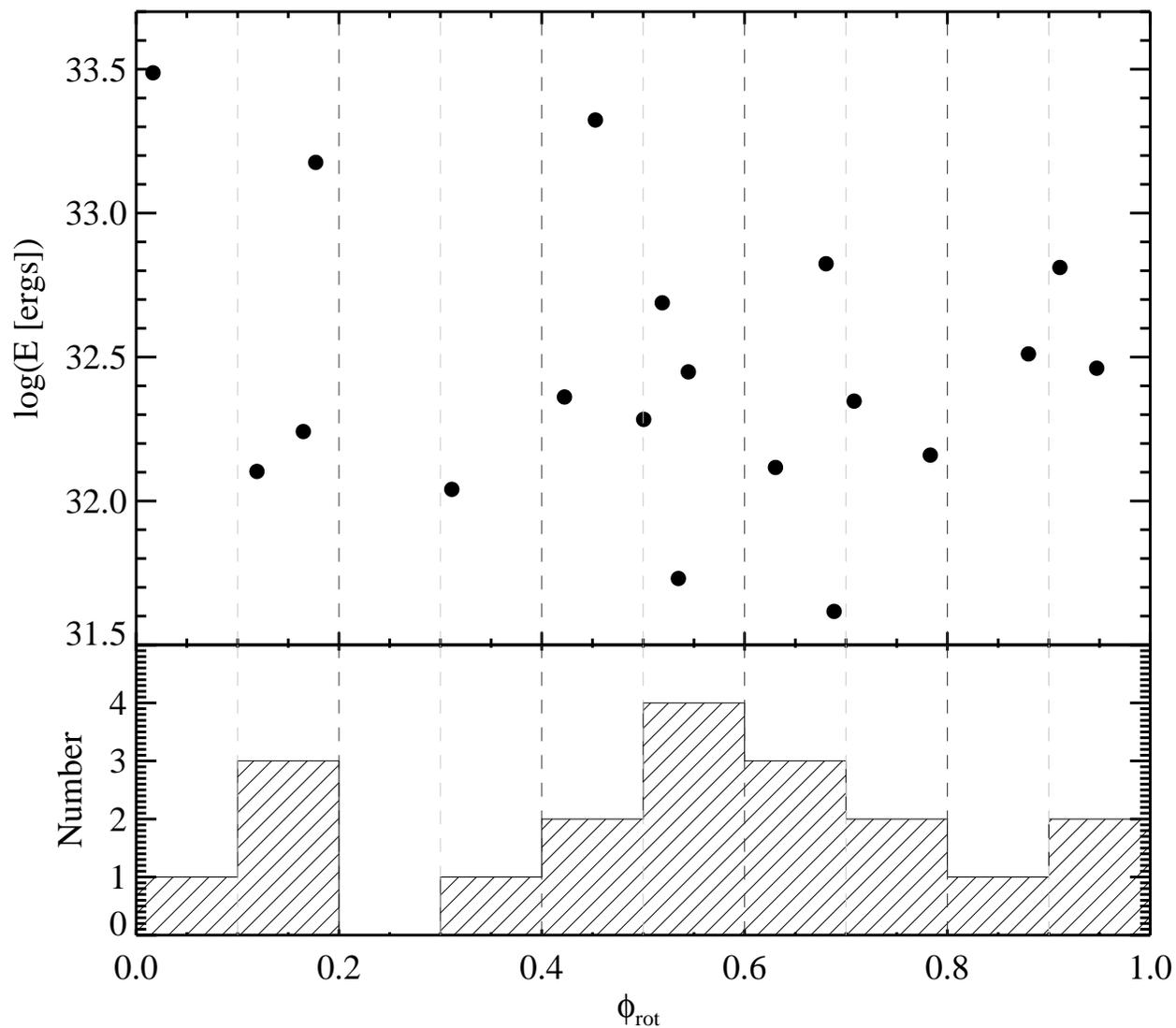}
\caption{\emph{Top:} Flare energy $E$ vs. rotational phase $\phi_{rot}$, where 0.5 is minimum light. 
\emph{Bottom: }Distribution of the number of flares in phase bins of 0.1.
We might expect to see more flares during minimum light, when the active region causing the modulation is most visible.  However, this is not apparent in our current observations.}
\label{fig:phase_corr}
\end{figure}

In the top panel of Figure \ref{fig:phase_corr}, we plot the flare energies against rotational phase, where a phase of 0.5 is the light curve minima.
The bottom panel shows the number of flares in each rotational phase bin of size 0.1.
The flares do not occur preferentially closer to minimum light, nor do the flares that occur near minimum light have larger energies.
This indicates that either 1) any difference between flare occurrence at minimum light and maximum light is not detectable in our small sample,
implying that the active region is large (or nearly pole on), such that even away from minimum light, flares that occur from the active region are visible; or 2) there are 
multiple active regions at a variety of longitudes that host flares, and while the rotational modulation is caused by the largest active region, flares may come from any of them.

Taken together, the distribution of flare wait times and the phase distribution of flare occurrence do not support a non-circumpolar single spot model, however, this result may also be due to low number statistics.
Future studies involving a much larger sample of flares from continuous monitoring, such as those that can be obtained with {\it Kepler} short-cadence ($\sim$ 1 minute) data,
will provide more robust statistics of these two measurements that may allow us to reconsider this model, as well as consider more complex models. 

%%%%%%%%%%%%%%%%%%%%%%%%%%%%%%%%%%%
%%%%%%%%%%%%%%%%%%%%%%%%%%%%%%%%%%%
\section{Conclusion}
\label{sec:conclusion}
%%%%%%%%%%%%%%%%%%%%%%%%%%%%%%%%%%%
%%%%%%%%%%%%%%%%%%%%%%%%%%%%%%%%%%%

We use over 8,500 individual {\it MOST} observations of AD Leo to find a rotational period of 2.23$^{+0.36}_{-0.27}$ dy.
Using the automated, iterative FIA of \citet{Hilton2012}, we identify 24 flare candidates, of which 19 pass visual inspection.
The FFD derived from these flares is consistent with previous studies, although it suffers from low number statistics and the observations 
were obtained with a much broader and redder filter.

The continuous monitoring provided by {\it MOST} allowed for a previously-unattainable wait-time analysis.  
Although the long waiting period leading up to the most energetic flare 
in this study suggests the possibility of a relationship between flare delay times and the occurrence of large, high-energy flares,
no meaningful relationship is apparent. 
Continuous monitoring and high precision photometry also allowed for the investigation of flare occurrence with rotational phase.  
We find no evidence of a relationship, although low-number statistics prevent us from ruling out the existence of such a relationship.

We compare these observational results to those expected from a simple model of magnetic energy storage and release.
In the model, magnetic energy is produced at a constant rate and stored in a single large magnetic spot, which acts as a reservoir. 
When the reservoir fills, magnetic energy is released in the form of flares.
The model predicts there to be a relationship between the wait-time of flares and flare energy, as well as between flare occurrence and
rotational phase. 
The model predictions are not supported by the observations.
This discrepancy could be explained if there are multiple reservoirs of magnetic energy or if the reservoir is not required to be full for a flare to occur.
One possible scenario, supported by \citet{Morin2008}, is of a large magnetic spot viewed nearly pole on.
This scenario predicts there to be a weak or nonexistent relationship between rotational phase and flare occurrence, but still predicts
there to be a relationship between flare energy and wait-time.

Further studies with longer monitoring times and enhanced photometric precision are necessary.
The \emph{Kepler} mission, in its short-cadence mode, allows for additional studies, the results of which will be presented in future papers.

\section{Acknowledgments}
We gratefully acknowledge support from NSF grant AST 08-07205, the Graduate Opportunities \& Minority Achievement Program, and from the {\it MOST} Consortium.
We thank James R. A. Davenport for useful discussions that have improved the paper.

%\bibliography{/Users/erichilton/Documents/hilton.bib}
\bibliography{most_refs.bib}

%%%%%%%%%%%%%%%%%%%%%%%%%%%%%%%%%%%%%%%%%%%%%%%%%%%%%%%%%%%%
%%%%%%%%%%%%%%%%%%%%%%%%%%%%%%%%%%%%%%%%%%%%%%%%%%%%%%%%%%%%
%figures
%%%%%%%%%%%%%%%%%%%%%%%%%%%%%%%%%%%%%%%%%%%%%%%%%%%%%%%%%%%%
%%%%%%%%%%%%%%%%%%%%%%%%%%%%%%%%%%%%%%%%%%%%%%%%%%%%%%%%%%%%

\end{document}